\documentclass[11pt]{article}
\usepackage{amssymb}
\usepackage{epsfig}
\usepackage{epsf}
\usepackage{here}
\usepackage{cite}
\usepackage{rotating}
\usepackage{array}
\begin{document}

\newcommand{\s}          {\mbox{$\rm{\sqrt{s}}$}}
\newcommand{\nsamp}      {\mbox{$\rm{\sqrt{N_{samples}}}$}}
\newcommand{\lnln}       {\mbox{$\rm{W^{+}W^{-}\to \ell \nu \ell' \nu '}$}}
\newcommand{\qqln}       {\mbox{$\rm{W^{+}W^{-}\to q \bar{q} \ell \nu }$}}
\newcommand{\qqqq}       {\mbox{$\rm{W^{+}W^{-}\to q \bar{q}  q \bar{q} }$}}
\newcommand{\GeV}        {\mbox{$\mathrm{GeV}$}}
\newcommand{\MeV}        {\mbox{$\mathrm{MeV}$}}
\newcommand{\pbinv}      {\mbox{$\mathrm{pb}^{-1}$}}
\newcommand{\GeVcc}      {\mbox{$\mathrm{GeV}\!/\!c^2$}}
\newcommand{\ww}         {\mbox{$\mathrm{W^{+}W^{-}}$}}
\newcommand{\mw}         {\mbox{$\rm{M_{W}}$}}
\newcommand{\mz}         {\mbox{$\rm{M_{Z}}$}}
\newcommand{\El}         {\mbox{$\rm{E_{\ell}}$}}
\newcommand{\Elmax}      {\mbox{$\rm{E^{max}_{\ell}}$}}
\newcommand{\Elmin}      {\mbox{$\rm{E^{min}_{\ell}}$}}
\newcommand{\ee}         {\mbox{$\rm{e^{+}e^{-}}$}}
\newcommand{\mumu}       {\mbox{$\rm{\mu^{+}\mu^{-}}$}}
\newcommand{\tautau}     {\mbox{$\rm{\tau^{+}\tau^{-}}$}}
\newcommand{\nunu}       {\mbox{$\rm{\nu\nu}$}}
\newcommand{\ZZ}         {\mbox{$\rm{ZZ}$}}
\newcommand{\Zee}        {\mbox{$\rm{Zee}$}}
\newcommand{\Wenu}       {\mbox{$\rm{We\nu}$}}

\newcommand{\etal} {\mbox{{\it et al.}}}
\newcommand{\CPC} {Comp. Phys. Comm.}
\newcommand{\PLB}[3]  {Phys.\ Lett.\ \textbf{B#1} (#2) #3}
\newcommand{\OPALColl}    {OPAL Collab.}
\newcommand{\ALEPHColl}    {ALEPH Collab.}
\newcommand{\EPJ} {Eur.~Phys.~J.} 
\newcommand{\PN} {Physics note} 
\newcommand{\NIM} {Nucl.~Instr.\ Meth.}
\newcommand{\Opal}       {\mbox{O{\sc pal}}}
\newcommand{\Aleph}      {\mbox{A{\sc leph}}}
\newcommand{\Delphi}     {\mbox{D{\sc elphi}}}
\newcommand{\Ltres}      {\mbox{L3}}
\newcommand{\LepII}      {\mbox{LEP2}}
\newcommand{\LepI}       {\mbox{LEP1}}
\newcommand{\Lep}        {\mbox{LEP}}
\newcommand{\WW}         {\mbox{WW}}
\newcommand{\KoralW}     {\mbox{K{\sc oral}W}}
\newcommand{\KoralZ}     {\mbox{K{\sc oral}Z}}
\newcommand{\Bhwide}     {\mbox{B{\sc hwide}}}
\newcommand{\Pythia}     {\mbox{P{\sc ythia}}}
\newcommand{\Kandy}     {\mbox{Y{\sc fsww3}}}
\newcommand{\Radcor}     {\mbox{R{\sc adcor}}}
\newcommand{\excalibur}  {\mbox{E{\sc xcalibur}}}
\newcommand{\grace}      {\mbox{{\tt grc4f}}}
\newcommand{\vermaseren} {\mbox{V{\sc ermaseren}}}
\newcommand{\herwig    } {\mbox{H{\sc erwig}}}

\begin{titlepage}
 \begin{center}{\large EUROPEAN ORGANIZATION FOR NUCLEAR RESEARCH 
 }\end{center}\bigskip
\begin{flushright}

  CERN-EP-2002-022\\ 11 March 2002

\end{flushright}
\bigskip\bigskip\bigskip\bigskip\bigskip
\begin{center}
 \huge\bf 
      Measurement of the Mass of the W Boson in $\mathbf{e^{+}e^{-}}$ 
      Collisions using the Fully Leptonic Channel
\end{center}
\bigskip\bigskip
\begin{center}{\LARGE The OPAL Collaboration}\end{center}
\bigskip\bigskip\bigskip
\begin{center}{\large
  Abstract}\end{center}
{A novel method of determining the mass of the W boson in the \lnln\ channel 
is presented and applied to
667 \pbinv\ of data recorded at center-of-mass energies in the range
183--207~\GeV\ with the \Opal\ detector 
at \Lep. The measured energies of charged leptons and the results of a new
procedure based on an
approximate kinematic reconstruction of the events are combined to give:
\begin{center}{$\rm{\mw\ = 80.41\pm 0.41\pm 0.13~\GeV,}$}
\end{center}
where the first error is statistical and the second is systematic.
The systematic 
error is dominated by the uncertainty on the lepton energy,
which is calibrated using data, and the parameterization of the
variables used in the fitting, which is obtained using Monte
Carlo events. Both of these are limited by statistics.
} 
\bigskip\bigskip\bigskip\bigskip
\bigskip\bigskip
\begin{center}{\large
(To be submitted to Euro. Phys. J.)
}\end{center}
\end{titlepage}

\begin{center}{\Large        The OPAL Collaboration
}\end{center}\bigskip
\begin{center}{
G.\thinspace Abbiendi$^{  2}$,
C.\thinspace Ainsley$^{  5}$,
P.F.\thinspace {\AA}kesson$^{  3}$,
G.\thinspace Alexander$^{ 22}$,
J.\thinspace Allison$^{ 16}$,
G.\thinspace Anagnostou$^{  1}$,
K.J.\thinspace Anderson$^{  9}$,
S.\thinspace Asai$^{ 23}$,
D.\thinspace Axen$^{ 27}$,
G.\thinspace Azuelos$^{ 18,  a}$
% ...
}\end{center}\bigskip
\bigskip
$^{  1}$School of Physics and Astronomy, University of Birmingham,
Birmingham B15 2TT, UK
\newline
$^{  2}$Dipartimento di Fisica dell' Universit\`a di Bologna and INFN,
I-40126 Bologna, Italy
\newline
$^{  3}$Physikalisches Institut, Universit\"at Bonn,
D-53115 Bonn, Germany
\newline
% ...
$^{  5}$Cavendish Laboratory, Cambridge CB3 0HE, UK
\newline
% ...
$^{  9}$Enrico Fermi Institute and Department of Physics,
University of Chicago, Chicago IL 60637, USA
\newline
% ...
$^{ 13}$Queen Mary and Westfield College, University of London,
London E1 4NS, UK
\newline
% ...
$^{ 16}$Department of Physics, Schuster Laboratory, The University,
Manchester M13 9PL, UK
\newline
% ...
$^{ 18}$Laboratoire de Physique Nucl\'eaire, Universit\'e de Montr\'eal,
Montr\'eal, Quebec H3C 3J7, Canada
\newline
% ...
$^{ 22}$Department of Physics and Astronomy, Tel Aviv University,
Tel Aviv 69978, Israel
\newline
$^{ 23}$International Centre for Elementary Particle Physics and
Department of Physics, University of Tokyo, Tokyo 113-0033, and
Kobe University, Kobe 657-8501, Japan
\newline
% ...
$^{ 27}$University of British Columbia, Department of Physics,
Vancouver BC V6T 1Z1, Canada
\newline
% ...
$^{  a}$ and at TRIUMF, Vancouver, Canada V6T 2A3
\newline
% ...
\medskip
\bigskip\bigskip\bigskip
\newpage
\section{Introduction}

At \LepII\ energies, 
four-fermion final states with two charged leptons and two neutrinos 
are dominated
by the pair production of W bosons. The probability of both W bosons
decaying into leptons is about
10\%. These events are characterized by an acoplanar pair of
charged leptons in the
final state \cite{bib:YB}. Systematic uncertainties due to colour 
reconnection, Bose-Einstein correlations and hadronization modeling,
 which plague the W mass measurement in other channels, 
are absent. The full reconstruction of the \lnln\ events 
is not possible due to the presence of at least two unobserved neutrinos 
in the final state. Therefore the W mass is determined from the energy 
spectrum of the charged leptons and from an approximate reconstruction 
of the event referred to as the pseudo-mass method.

This paper presents a measurement of the W mass \mw\ in the fully leptonic
channel with the \Opal\ detector, using an unbinned maximum likelihood fit. 
The data sample has an integrated luminosity of 667 \pbinv\ recorded at
center-of-mass energies between 183 \GeV\ and 207 \GeV.

A general description of the
reconstruction of \lnln\ events can be found in \cite{bib:desy}.
The end-points of the leptonic energy spectrum in \lnln\ events depend
on the W mass. Neglecting the masses of the charged leptons 
and the finite width of the W boson, the energy \El\ of the charged leptons 
can be written in terms of \mw\ as
\footnote{The convention c=1 is used throughout this paper.}:
\begin{equation}
\mathrm{E_{\ell}= \frac{\sqrt{\it{s}}}{4} + \cos \theta ^{*}_{\ell}
\sqrt{\frac{\it{s}}{16}-\frac {M^{2}_{W}}{4}}}
\end{equation}
\noindent where $\it{s}$ is the square of the center-of-mass energy and 
$\rm{\theta^{*}_{\ell}}$ is the angle 
between the lepton direction measured in the W rest frame and 
the direction of the W in the laboratory frame. 
The latter is not known, so the W mass is determined mainly by the endpoints
of the distribution, which correspond to $\rm{\cos \theta^{*}_{\ell} = \pm1}$.
In practice, however, the end-points of the distribution are 
smeared considerably by the width of the W boson, by initial state radiation
and by the detector resolution. These effects weaken the
sensitivity of this variable. 

In order to include the information from the angle between the two
leptons and the correlation between their energies, a second observable based
on an approximate kinematic reconstruction of the event --- the so-called
pseudo-mass --- is used in the analysis. 
The complete reconstruction of the event would require the determination 
of the four-momenta of both charged leptons and the two neutrinos, 
12 quantities in total assuming the masses are known. The four-momenta 
of the two charged leptons are measured; therefore, the full 
reconstruction depends on the determination of the momenta of the two 
neutrinos. Assuming four-momentum 
conservation and equal masses for the two W bosons, five more constraints can be obtained. 
An additional arbitrary constraint is imposed to allow the ``reconstruction''
of the event. By assuming that both neutrinos are in the same plane as the
charged leptons, the kinematics can be solved to yield a W \emph{pseudo-mass},
which is quite sensitive to the true W mass. 
Due to a twofold ambiguity, two solutions are found for this variable:

\begin{eqnarray}
M^{2}_{\pm} & = & \frac{2}{(\mathbf{p_{\ell'}}+\mathbf{p_{\ell}})^{2}}
\Big((P~\mathbf{p_{\ell'}}-Q~\mathbf{p_{\ell}})(\mathbf{p_{\ell'}}+\mathbf{p_{\ell}})\\
                    &   & \pm \sqrt{(\mathbf{p_{\ell}}\times 
\mathbf{p_{\ell'}})^{2}[(\mathbf{p_{\ell'}}+\mathbf{p_{\ell}})^{2}(E_{\mathrm{beam}}-E_{\ell})^{2}-(P+Q)^{2}]}\Big),
\nonumber
\end{eqnarray}

\noindent where
\begin{displaymath}
P = E_{\mathrm{beam}}E_{\ell}-E^{2}_{\ell}+\frac{1}{2}m^{2}_{\ell},\qquad
Q = -E_{\mathrm{beam}}E_{\ell'}-\mathbf{p_{\ell'}}\cdot 
\mathbf{p_{\ell}}+\frac{1}{2}m^{2}_{\ell'},
\end{displaymath}
\noindent $E_{\rm{beam}}$ is the beam energy, 
$E_{\ell},~E_{\ell'}$ are the energies of the 
two charged leptons, $m_{\ell},~m_{\ell'}$ are their masses 
and $\mathbf{p_{\ell}},~\mathbf{p_{\ell'}}$ 
are their three-momenta. In the present analysis the charged leptons are 
considered to be massless particles because the values of the masses
are very small in comparison with their energies.
The sensitivity of both solutions was studied with Monte Carlo 
simulations, which 
showed that only the larger solution, $M_{+}$, is sensitive to \mw.
The distribution of this solution shows
an edge where the value of the pseudo-mass is close to  
the true W mass. 

The pseudo-mass is used  here for the first time in determining the
W mass in leptonic W-pair decays. It is shown that the correlation
between the mass determination from the single lepton energy spectrum and 
from the pseudo-mass is small, so that the mass measurement 
can be improved by combining the two methods. For the measurement 
presented in this paper the total
error is dominated by the data statistical error.
The largest contributions to the systematic uncertainty 
in this analysis are due to the uncertainties on the lepton energy 
which are determined with data and scale with the available statistics.
In future experiments with larger 
statistics it should be possible to reduce both the statistical and systematic
uncertainty significantly.

\section{Detector Description and Event Selection}

A detailed description of the \Opal\ detector can be found 
in \cite{bib:detect}. The data sample used for this analysis 
corresponds to an accepted integrated luminosity of 667 $\rm{pb^{-1}}$, 
evaluated using small angle Bhabha scattering events observed in 
the silicon tungsten forward calorimeter\cite{bib:lumi}. 

The selection of the \lnln\ events and the identification of 
electrons and muons are described in \cite{bib:graham,bib:graham2}. 
For this analysis events with two charged leptons are required.
In view of the experimental problems associated with 
reconstructing $\rm{\tau}$-decays, the present analysis is restricted to
events which contain at least one W decaying 
into an electron or a muon for the leptonic energy method (section 3.1.1) 
and events in which neither of the two W bosons decays into a tau
 for the pseudo-mass method (section 3.1.2). Table \ref{t:eff} summarizes
the signal efficiencies and purities \cite{bib:cross-sec}. 
The dominant background sources are $\rm{We\nu, ~ZZ}$ and dilepton events. 
These sources comprise 40\%, 30\% and 15\% 
of the total background cross-section 
for the \lnln\ events \cite{bib:cross-sec}. 
A total of 1101 events for the whole range of center-of-mass 
energies are observed in the data. Table \ref{t:events} summarizes the 
observed numbers of events for each center-of-mass energy together with the 
corresponding integrated luminosities. 

\subsection{Monte Carlo Event Generators}

At each center-of-mass energy, samples of signal events were generated with
five different W masses: 79.33 \GeV, 79.83 \GeV, 80.33 \GeV, 80.83 \GeV\ and
81.33 \GeV. In the range of 183--189 \GeV, samples corresponding to the
central value of 80.33 \GeV\ were generated according to
the CC03 \footnote{The leading order \ww\ production
diagrams, i.e.~the $t$-channel $\rm{\nu_{\mathrm{e}}}$ exchange and the 
$s$-channel $\rm{Z^{0}/\gamma}$ exchange, are referred to as CC03, 
following the notation of \cite{bib:channel}.} diagrams 
with \KoralW\  \cite{bib:koralw}, $\rm{e^{+}e^{-}\to ZZ~and~Zee}$ background events 
were generated with \Pythia\ 
\cite{bib:pythia} and $\rm{We\nu}$ with
\KoralW\ \cite{bib:koralw} . 
For the other W masses and for center-of-mass energies in the range of
192--207 \GeV\ \KoralW\ was used both for the signal and for these
four-fermion backgrounds. Other backgrounds were simulated by \Radcor\ 
\cite{bib:radcor} for multi-photon final states, \KoralZ\ \cite{bib:koralz} 
and \Bhwide\ \cite{bib:bhwide} for dilepton final states and
two-photon events were simulated with \vermaseren\ \cite{bib:vermaseren} 
and \herwig \cite{bib:herwig}. Finally \Kandy\ \cite{bib:kandy}
samples were used to perform 
systematic studies related to photon radiation in \ww\ events.

\subsection{Classification of Events}\label{section:class}

\indent The momentum resolutions obtainable with \Opal\ for 
electrons and muons differ significantly at high energies.
For electrons, the preferred measurement uses the electromagnetic 
calorimeter energy information which has a resolution of
approximately 3\% at 45 \GeV, whereas for muons 
at the same energy, the charged particle momentum determined in 
the central tracker has a resolution of 
approximately 8\% \cite{bib:opal-detector}. Therefore, to maximize 
the sensitivity of the measurement, electrons and muons are 
treated separately by defining different classes of events for
the leptonic energy and the pseudo-mass analyses, depending 
upon the identified lepton flavours. Identified 
taus are rejected since their energy cannot be determined.

In the case of the leptonic energy the information of both charged leptons
is used independently. For this variable two different classes are defined: 
the first class contains leptons identified as electrons
and the second class contains leptons identified as muons.
To increase the number of leptons used in the analysis, the higher
energy identified electrons or muons in events selected as either 
$e\nu_{e}\tau\nu_{\tau}$ or $\mu\nu_{\mu}\tau\nu_{\tau}$ are also used.

For the pseudo-mass three classes of events are defined as follows:
\begin{enumerate}
\item Events consisting of two leptons identified as $e$. 
\item Events consisting of two leptons identified as $\mu$.  
\item Events consisting of two leptons identified as an $e$ and a $\mu$. 
\end{enumerate}

Table \ref{t:classi} shows the classification of events for the 
leptonic energy and the pseudo-mass. The number of leptons selected 
for the leptonic energy 
method, and the number of events selected for the pseudo-mass 
method are summarized in Table \ref{t:events}.

\section{Extraction of the W Mass}

\subsection{Unbinned Maximum Likelihood Method}

The extraction of the W mass was performed by
a simultaneous maximum likelihood fit to the leptonic energy and 
the pseudo-mass distributions using the data from center-of-mass energies from 
183 \GeV\ to 207 \GeV. The fit is performed with the following product of
likelihood functions:
\begin{equation}
\mathcal{L}_{T} = \prod_{k=1}^{8}\left(\prod_{i=1}^{2}\mathcal{L}_{LE}^{k}\times 
\prod_{i=1}^{3}\mathcal{L}_{PM}^{k}\right)
\end{equation}
\noindent where $k$ denotes the eight center-of-mass energies included
in the fit and $i$ runs over the two classes defined for the leptonic
energy ($LE$) and the three classes defined for the pseudo-mass ($PM$).

The leptonic energy and the pseudo-mass distributions
obtained from Monte Carlo simulations (including all background 
sources) are parameterized by appropriate analytic functions,
\begin{equation}\label{eq:equ1}
f = f(P_{1},\cdots ,P_{N};x), 
\end{equation}
\noindent which depend on $N$ parameters, $P_{i}$, and on the variable
$x$, which stands for either the leptonic energy or the pseudo-mass.
For each parameter a linear dependence on \mw\ is assumed:

\begin{equation}
P_{i} = b^{0}_{i} + b^{1}_{i}\times \mw
\end{equation}
The coefficients $b^{0}_{i}$ and $b^{1}_{i}$ are obtained 
by fitting the analytic functions to the leptonic energy and 
the pseudo-mass spectra generated at different W masses. 
The coefficient $b^{1}_{i}$ is included only 
if it is not compatible with zero
within one standard deviation 
(i.e.~its numerical value is larger than the
corresponding error); otherwise only a constant term is used. 
This parameterization is performed independently for 
each class defined for the sensitive variables. 
Details of the method are discussed in the following sections. 

\subsubsection{Parameterization of the Leptonic Energy}\label{section:par_en}

The leptonic energy spectrum
is fitted with a function $f_{T}$, which is the product
of two Fermi functions and a linear function: 
\begin{equation}
f_{T} = f_{1} \times f_{2} \times f_{3}
\end{equation}
\noindent with
\begin{equation}\label{eq:en_func}
f_{1} = \frac{1}{e^{-\frac{x-P_{1}}{P_{2}}} + 1}, \quad f_{2} = 
P_{5}(1+P_{6}\times x),
\quad f_{3} = \frac{1}{e^{\frac{x-P_{3}}{P_{4}}} + 1}.
\end{equation}
\noindent $P_{1}$ and $P_{3}$ correspond to the points of inflection
of the two Fermi functions, $P_{2}$ and $P_{4}$ to their widths,
$P_{5}$ is the constant term of the linear function and the
 slope is proportional to $P_{6}$. 

\begin{figure}[H]
\begin{center}
\epsfig{file=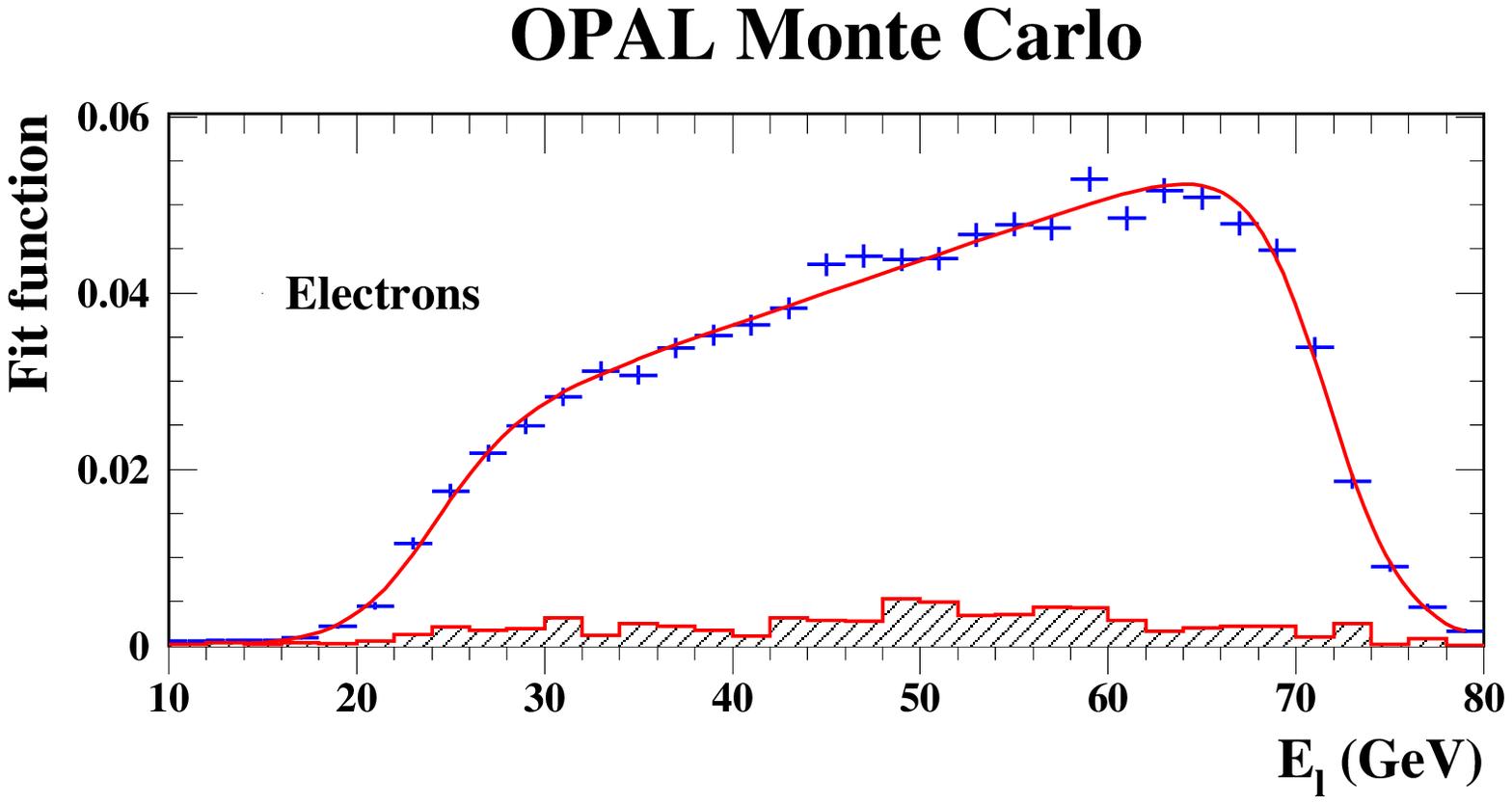,width=15cm}  
\caption{Fit to the leptonic energy 
distribution generated with \mw\ = 80.33 \GeV\ for \s~=~189~\GeV.
The crosses indicate Monte Carlo events and the shaded area shows the
background Monte Carlo. Only leptons tagged as electrons were used 
for this distribution. The fitted function has five free parameters and 
is a product of two Fermi functions and a linear function.}
\label{fig:pr353_01}
\end{center}
\end{figure}

The fit function depends therefore on six parameters. The overall normalization
of the spectra can be used to eliminate $P_{5}$. For each 
set of values of the $P_{1}$, $P_{2}$, $P_{3}$, $P_{4}$ 
and $P_{6}$ parameters, $P_{5}$ is calculated so that
the total function is normalized in the region taken to perform
the fit. The lower limit is 10~\GeV\ for all center-of-mass 
energies and the upper limit changes depending on the center-of-mass energy
from 80~\GeV\ for 183~\GeV\ to 100~\GeV\ for 207~\GeV. The regions at the end-points
of the leptonic energy spectrum are sensitive to \mw.
In terms of the parameterization these edges correspond to
$P_{1}$ and $P_{3}$. These are the only 
parameters which are expected to change with the W mass while
the other parameters, $P_{2}$, $P_{4}$ and  $P_{6}$, are
more sensitive to the W width and detector resolution effects.
This has been confirmed by studying the linear dependence of each parameter 
on \mw. Figure \ref{fig:pr353_01} shows an example of a fit to the
leptonic energy distribution. The spectrum is generated from a Monte Carlo 
sample with a W mass of 80.33~\GeV\ at a center-of-mass energy of 189~\GeV. 

\begin{figure}[H]
\begin{center}
\epsfig{file=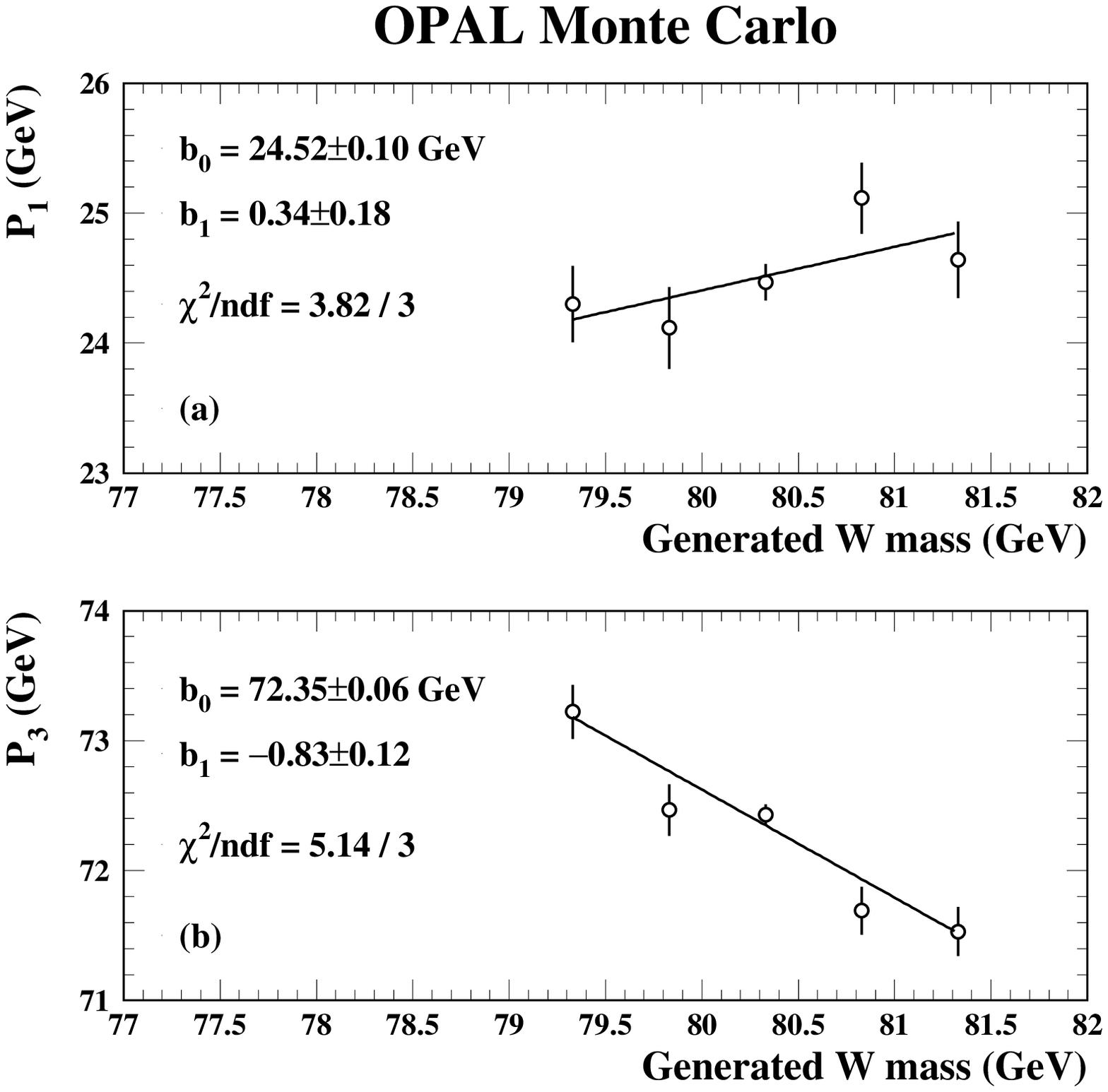,width=15cm}  
\caption{Linear fit ($P_{i} = b_{0}+b_{1}\times(\rm{M_{W}}-80.33)$)
of the coefficients (a) $P_{1}$ and (b) $P_{3}$
(points of inflection of the Fermi functions) 
at a center-of-mass energy of 189 \GeV.
The events chosen to perform this fit belong to the first class defined for the
leptonic energy, which contains electrons only. Similar studies are performed
for the second class and at all center-of-mass energies.}
\label{fig:pr353_02}
\end{center}
\end{figure}

A simultaneous fit of the parameters $P_{1}(\mw)$, $P_{2}$
and $P_{3}(\mw)$, $P_{4}$ and $P_{6}$
with the Monte Carlo samples for different values of \mw\ is performed. 
Parameters $P_{2}$, $P_{4}$ and $P_{6}$ are assumed to be 
independent of \mw\ and a common value 
is determined for each of these. $P_{1}$ and $P_{3}$
describe the \mw\ dependence and a separate value is determined for each 
generated \mw. Figure \ref{fig:pr353_02} shows the fitted parameters 
$P_{1}$ and $P_{3}$ and the linear dependence 
on \mw\ for the case of electrons.
These fits are performed independently for each class of events and each 
center-of-mass energy.

\subsubsection{Parameterization of the Pseudo-mass}\label{section:par_pm}

\begin{figure}[H]
\vspace{-3cm}
\begin{center}
\epsfig{file=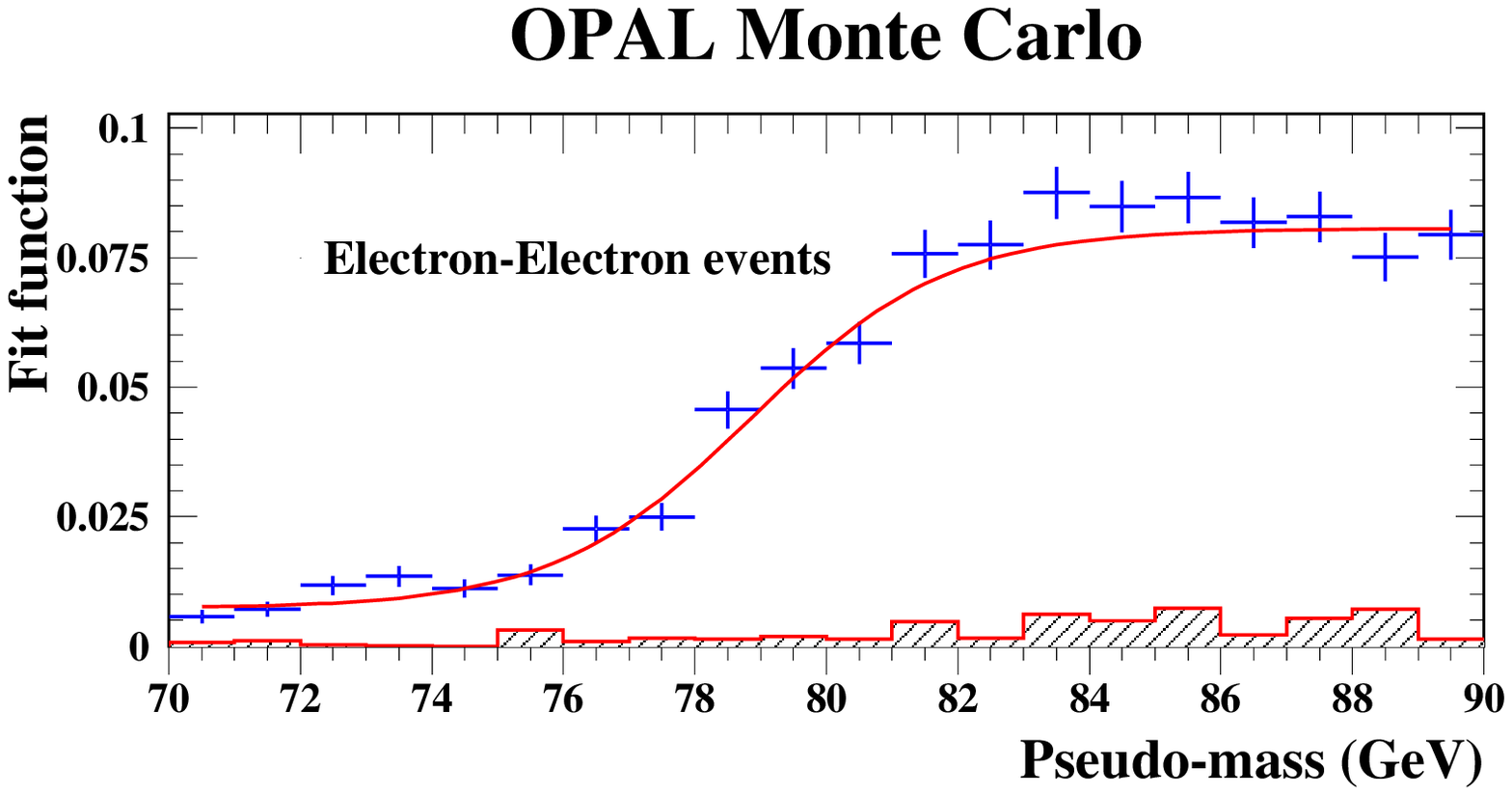,width=15cm}  
\caption{Fit to the pseudo-mass
distribution generated with \mw\ = 80.33 \GeV\ for \s~=~189~\GeV.
The crosses indicate simulated Monte Carlo events and the shaded area shows
the background Monte Carlo. Only events tagged as electron-electron were used. 
The fitted function has three free parameters and consists of a 
Fermi function plus a constant.}
\label{fig:pr353_03}
\end{center}
\end{figure}

\begin{figure}[H]
\begin{center}
\epsfig{file=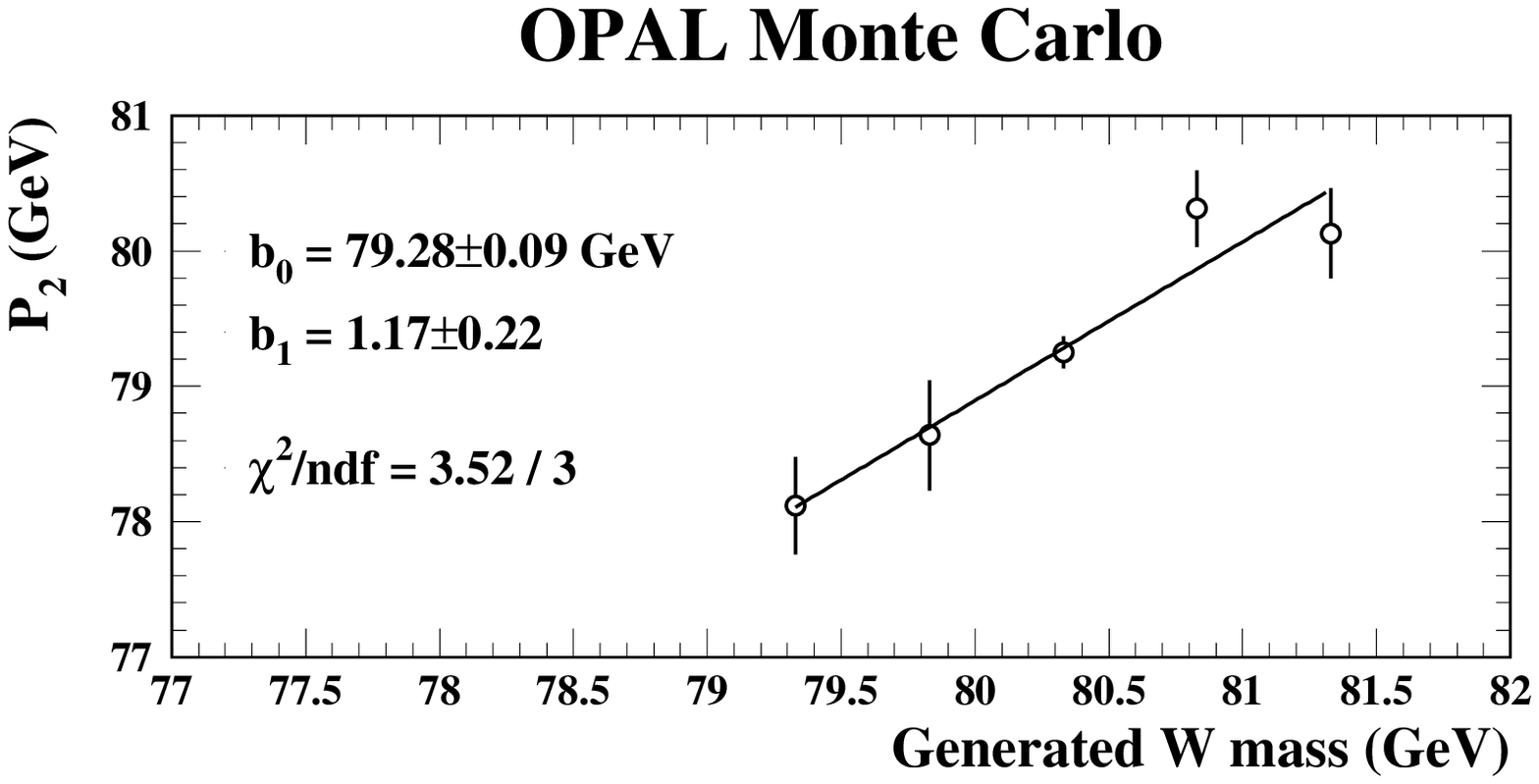,width=15cm}  
\caption{Linear fit ($P_{2} = b_{0}+b_{1}\times(\mathrm{M_{W}}-80.33)$) 
of the coefficient $P_{2}$ (point of inflection of the Fermi function) 
to \mw\ at a center-of-mass energy of 189~\GeV.
The events chosen to perform this fit belong to the first class defined for the
pseudo-mass, which contains electrons only. Similar studies are performed
for the second and third classes and at all center-of-mass energies.}
\label{fig:pr353_04}
\end{center}
\end{figure}

The analytic function chosen to fit the pseudo-mass spectra is the sum
of a Fermi function and a constant function, 
with four free parameters:

\begin{equation}\label{eq:fermi}
f = P_{1}\Bigg (\frac{1}{e^{-\frac{x-P_{2}}{P_{3}}} + 1} +P_{4}\Bigg).
\end{equation}
\noindent $P_{1}$, $P_{2}$ and $P_{3}$ are proportional to
the amplitude of the curve, to the point of inflection of the slope and to the width
respectively and $P_{4}$ is a constant term. As with the 
lepton energy, one parameter can be eliminated due to the constraint of the
overall normalization. This procedure is used to eliminate 
the parameter $P_{1}$. For each set of values of the 
$P_{2}$, $P_{3}$, $P_{4}$ parameters, $P_{1}$ is determined to
normalize the function within the fit range of 70~\GeV\ to 90~\GeV, 
which is the same for all center-of-mass energies. In terms 
of the above parameterization, the edge of the pseudo-mass distribution
corresponds to $P_{2}$. This is the
only parameter which is expected to show a clear dependence on \mw. 
Figure \ref{fig:pr353_03} shows an example of the function fitted to a 
simulated pseudo-mass distribution. 
Similar fits are performed for each class of events 
at all center-of-mass energies considered in the analysis. 
The edge around the generated W mass can be observed in the figure.
 The individual linear dependence of each parameter on \mw\ was studied in
the same way as for the leptonic energy.
In this case, a simultaneous fit of the parameters $P_{2}(\mw)$, $P_{3}$
and $P_{4}$ with the Monte Carlo samples for the different values of
\mw\ was performed. $P_{3}$ and
$P_{4}$ are found to be independent of \mw; therefore a common value 
was determined for each. $P_{2}$ describes the \mw\ dependence and an
independent value was determined for each generated \mw. 
Figure \ref{fig:pr353_04} shows the fitted values of the $P_{2}$ parameter 
and its linear dependence on \mw\ in the case of electron-electron events.

\subsection{Monte Carlo Studies}

The correlation between the fitted masses for the leptonic energy
and for the pseudo-mass has been determined to be ($11\pm1$)\%, considering
all center-of-mass energies together. 
This was evaluated by fitting the W mass separately for the leptonic
energy and the pseudo-mass with an ensemble of 90 independent
data-sized Monte Carlo subsamples generated with a W mass of 80.33 \GeV. 

\begin{figure}[H]
\begin{center}
\epsfig{file=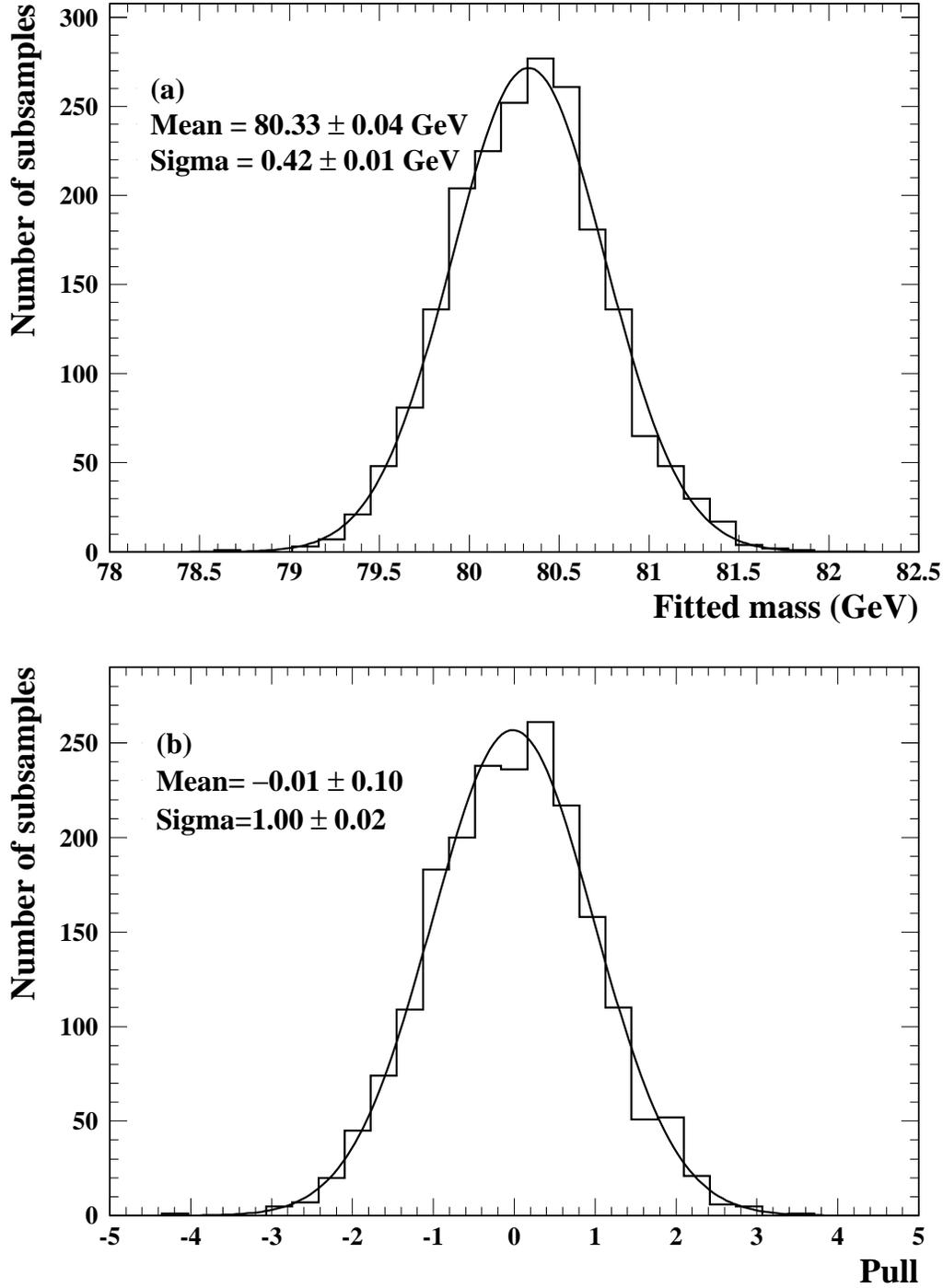,width=15cm}  
\caption{(a) Fitted mass distribution from Monte Carlo for the simultaneous
fit of the leptonic energy and the pseudo-mass at all center-of-mass 
energies. The test samples are generated at \mw\ = 80.33 \GeV. (b) The 
corresponding pull distribution. The errors have been rescaled by a 
factor 1.11 to take into account the correlation between the 
leptonic energy and the pseudo-mass.}
\label{fig:pr353_05}
\end{center}
\end{figure}

The simultaneous fit method has been tested with 2000 data-size experiments
derived by resampling from Monte Carlo simulations. 
The events of these subsamples were picked at
random from the full Monte Carlo sample, whose signal part
(generated at \mw\ = 80.33 \GeV) corresponded to 90 times the data statistics. 
Multiple draws of the same events were allowed. Figure \ref{fig:pr353_05}(a) 
shows the results of the resampling tests fitted with a Gaussian function. 
The pull distribution is shown in figure \ref{fig:pr353_05}(b).
To compensate for the underestimation of the statistical error by the 
common fit due to the correlation between 
the leptonic energy and the pseudo-mass, the errors of the fit used in the 
pull distribution have been increased by a factor 1.11.

\subsubsection{Bias Test}

\begin{figure}[H]
\begin{center}
\epsfig{file=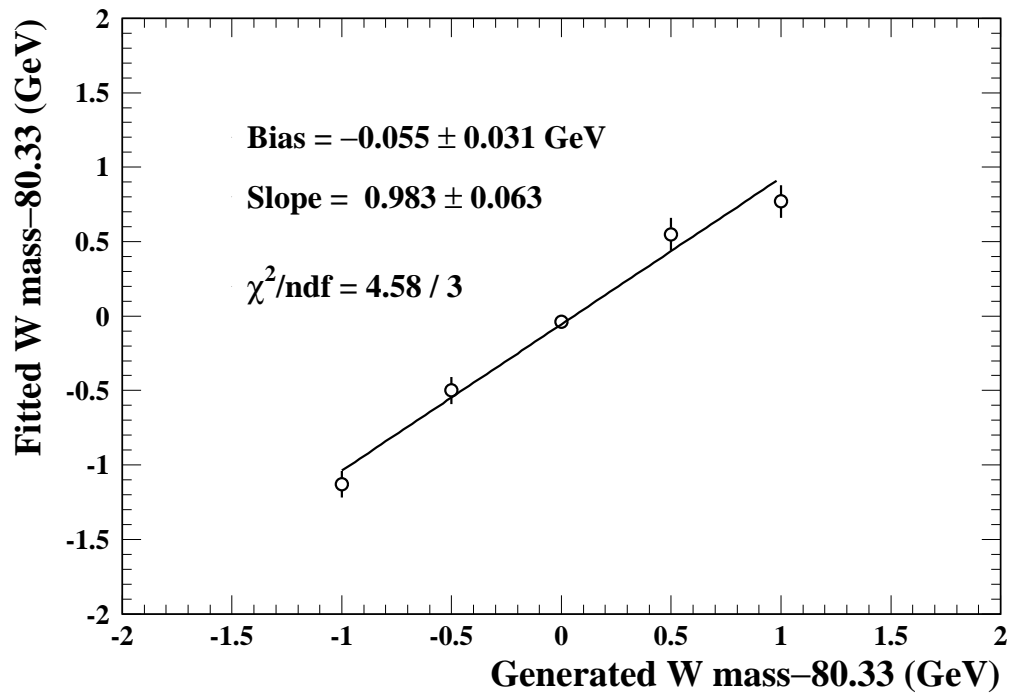,width=15cm}  
\caption{Fitted versus generated W mass, showing the linearity 
of the unbinned method. The central
value of 80.33 \GeV\ is subtracted from all the masses.}
\label{fig:pr353_06}
\end{center}
\end{figure}

The method presented to extract the W mass is not expected to show any bias. 
Possible biases due to detector effects are eliminated since these are 
simulated with Monte Carlo methods. 
Biases related to the selection of the analytic
functions which fit the leptonic energy and the pseudo-mass are negligible
as long as the functions fit the variables properly. By using Monte Carlo
samples generated with \mw\ = 79.33, 79.83, 80.33, 80.83 and 81.33 \GeV, 
the bias and the linearity can be determined. 
First, the individual relations between 
the generated and the fitted masses are studied independently 
for the leptonic energy and for the pseudo-mass. 
The distributions show slopes and biases compatible with one and zero
respectively, $\rm{0.958\pm 0.087}$ and $\rm{-0.061\pm 0.054~\GeV}$
for the leptonic energy and $\rm{0.952\pm 0.086}$ and 
$\rm{-0.031\pm 0.044~\GeV}$ for the pseudo-mass. Second, the relation between 
the generated and the fitted mass is studied for the simultaneous fit to both
the leptonic energy and the pseudo-mass. This relation 
is shown in Figure \ref{fig:pr353_06} and found to be linear in a region 
of $\pm$ 1 \GeV\ from the central value of 80.33 \GeV, with a slope of 
$\rm{0.983\pm 0.063}$ and a bias value of $\rm{-0.055\pm 0.031~\GeV}$. 
No corrections are included due to the slope or bias values since there is
no evidence of a bias. 

\section{Fit Results and Discussion}

\begin{figure}[H]
\begin{center}
\epsfig{file=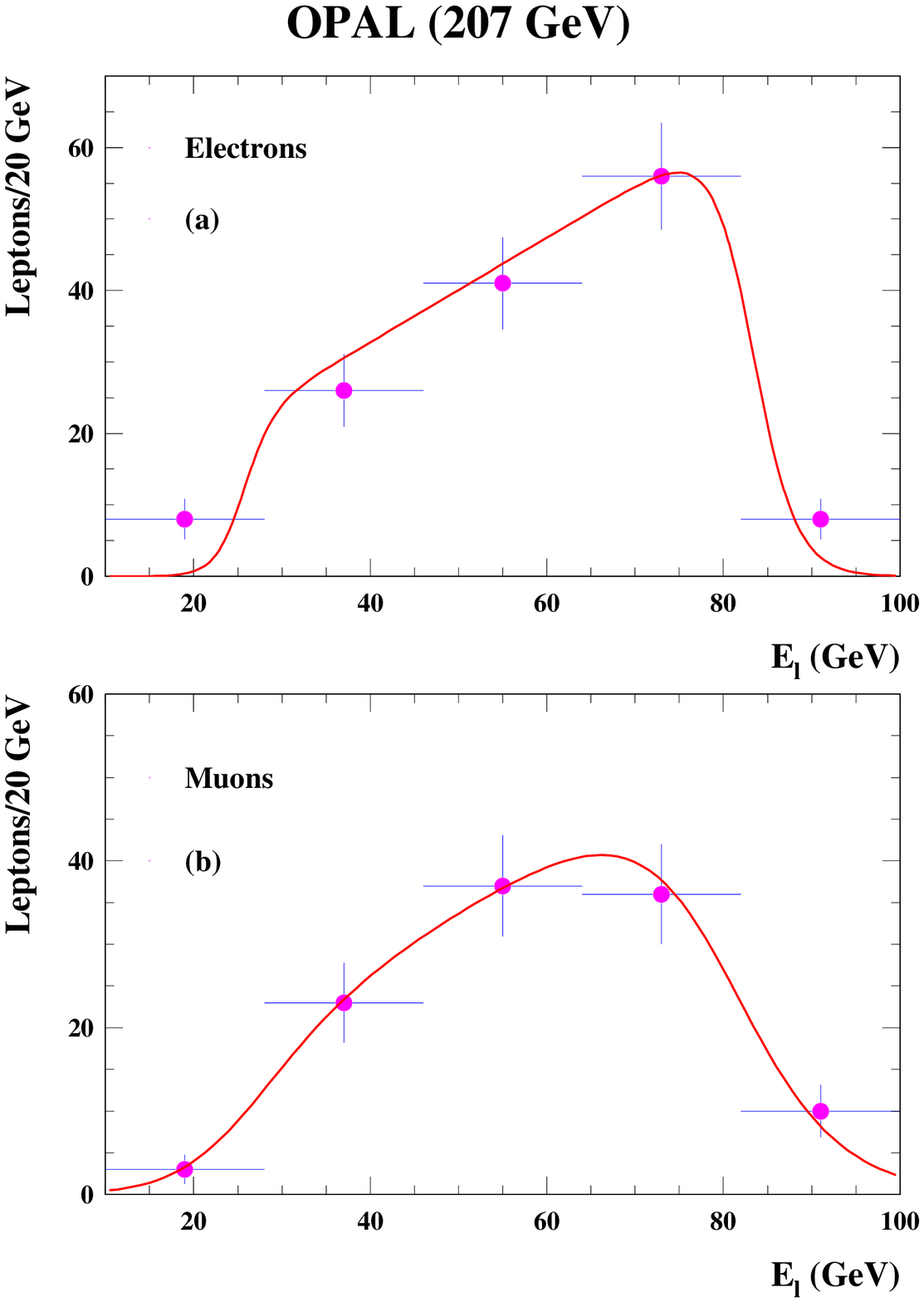,width=15cm}  
\vspace{-1truecm}
\caption{Comparison between the data and the 
fit functions for the leptonic energy at a center-of-mass energy 
of 207 \GeV. (a) electrons. (b) muons.} 
\label{fig:pr353_07}
\end{center}
\end{figure}

\begin{figure}[H]
\begin{center}
\epsfig{file=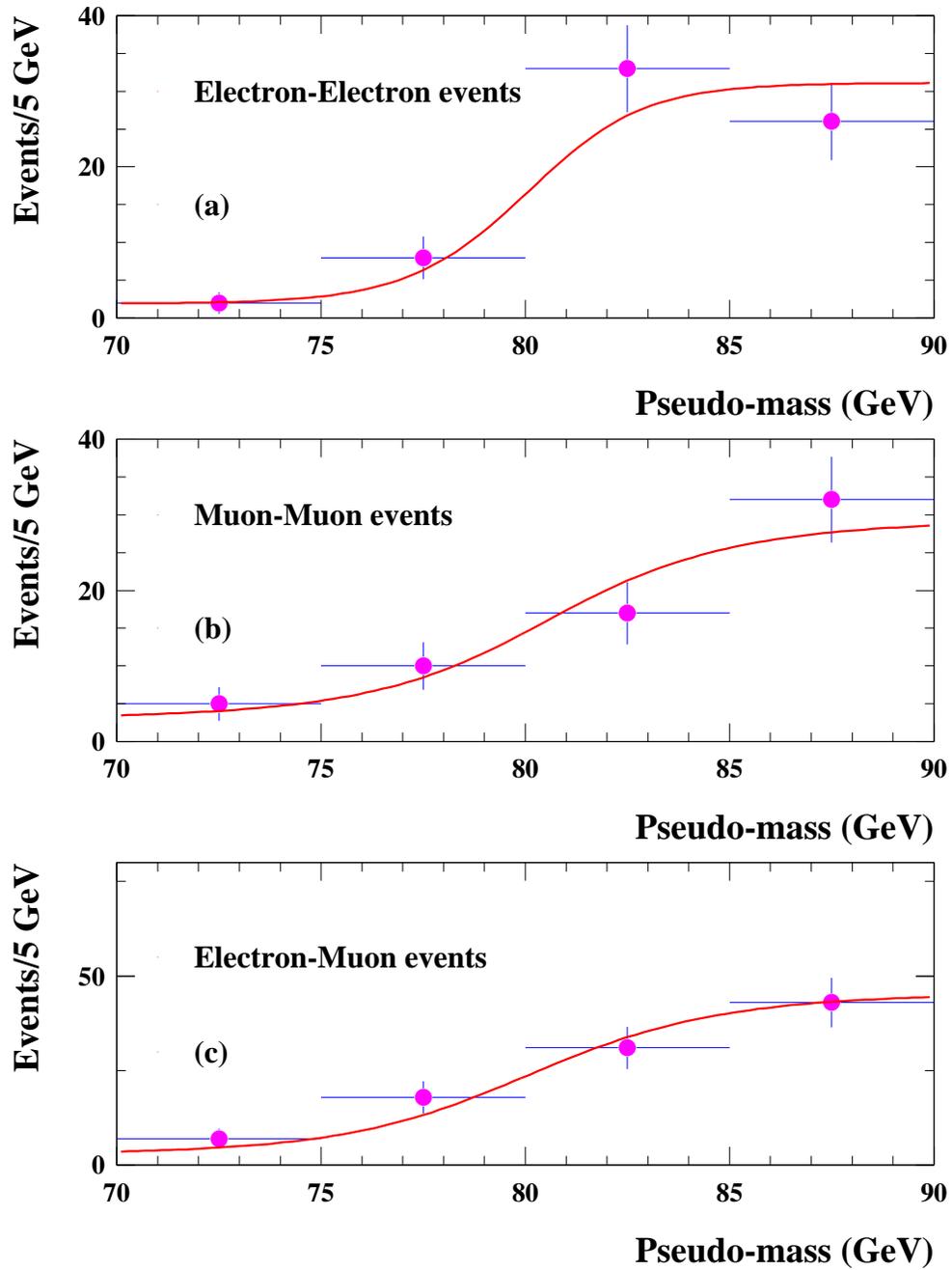,width=15cm}  
\vspace{-1truecm}
\caption{Comparison between the data and the 
fit function for the pseudo-mass at all center-of-mass energies.
(a) electron-electron events. (b) muon-muon events. (c) electron-muon
events.}
\label{fig:pr353_08}
\end{center}
\end{figure}

The simultaneous fit to the data distribution of
 the leptonic energy and the pseudo-mass, combining 
center-of-mass energies from 183 \GeV\ to 207 \GeV, 
gives the following result:
\begin{center}{$\rm{\mw\ = 80.41\pm 0.41~\GeV,}$}
\end{center}
where the quoted error is statistical only and has been scaled
 by a factor 1.11 to
take into account the correlation between the two variables. The W mass values
obtained separately are $\rm{\mw = 80.58\pm 0.52~\GeV}$ for the
leptonic energy and $\rm{\mw = 80.20\pm 0.61~\GeV}$ for the pseudo-mass. 
Figure \ref{fig:pr353_07} shows the comparison between the 
fit function and the data for the two classes defined
for the leptonic energy at a center-of-mass energy of 207 \GeV. 
The poorer resolution of the muon energy compared to the 
electron energy is visible in the fit functions. The regions around the 
end-points have sharper edges for electrons than for muons. 
Figure~\ref{fig:pr353_08} shows analogous results
for the three classes of events defined for the pseudo-mass at all
center-of-mass energies.  
The greater sensitivity of electron-electron events
 relative to muon-muon or electron-muon events can be observed
by comparing the fit functions around 80 \GeV, i.e. in the region sensitive
to \mw. 

\section{Systematic Checks and Uncertainties}

The study of the systematic uncertainties on the W mass is 
described in this section and summarized in Table \ref{t:sys_un}. 
The sources of systematic errors have been studied simultaneously for all 
center-of-mass energies. The total systematic error is calculated as a 
quadratic sum of the contributions listed below. 

\subsection{Beam Energy}

The average \Lep\ beam energy is currently known with a precision of
about $\pm$ 20 \MeV, varying slightly
for different center-of-mass energies between 183 \GeV\ and 207 \GeV
\cite{bib:lep-energy}. This leads to a systematic uncertainty of 11 \MeV. 
The RMS spread in the \Lep\ center-of-mass energy is around 240 \MeV,
varying slightly with the center-of-mass energy. 
The  systematic error due to the uncertainty of the beam energy
spread is 3 \MeV. 
In addition the data
at average center-of-mass energies of 205 \GeV\ and 207 \GeV\ were taken at
a number of discrete energy points in the range of 204.6--206.0 \GeV\ and
206.2--208.0 \GeV, respectively. This leads to a bias of 12 \MeV\ for 
which the fit result has been corrected.
The average difference between the electron and the
positron beam energies has negligible impact on the measurement of \mw.

\subsection{QED Corrections}

The systematic error associated with uncertainties in the 
modeling of the QED corrections is estimated by 
reweighting events generated with \Kandy, which has a more
complete treatment of $ {\cal O}(\alpha)$ QED corrections,
according to \KoralW\ probabilities.
The corresponding difference of 7~\MeV\ is taken as systematic error.

\subsection{Detector and Resolution Effects}

The effects of the detector calibrations and the deficiencies 
in the Monte Carlo simulation of the detector response are 
investigated by varying the observed leptonic energy scales in 
the electromagnetic calorimeter (for the electrons) and in 
the central tracker (for the muons) over reasonable ranges 
\cite{bib:mass_189}.
The ranges used for the systematic variations depend on the polar angle and
are determined from detailed comparisons of data and Monte Carlo utilizing
both data recorded at 189--207 \GeV\ and data collected at
\s\ $\approx$ \mz\ during 1998--2001. Due to differences 
observed in the resolution  between the data
and the Monte Carlo simulation during these years, correction factors for
electron and muon energy are included in the reference samples. The difference 
between the measured and the generated energy is scaled by a factor 
1.07 for the electromagnetic calorimeter and 1.06 for the central tracker. 

From the systematic studies of the lepton energy scale, the electromagnetic 
calorimeter energy scales and the central tracker scale
are both known to 0.3\%. These scale uncertainties dominate the 
detector related systematics in the \lnln\ channel. For muons the 
scale uncertainty is  32~\MeV.  
In the case of electrons, 
the scale errors are  expected to be uncorrelated between 
the barrel and endcap regions and therefore the analysis for electrons 
was performed separately in both regions and combined 
in quadrature, resulting in an uncertainty of 85~\MeV. The systematic 
uncertainty due to the non-linearity in the electromagnetic calorimeter
 and the central tracker is taken 
into account by varying the energy using a factor which changes linearly from
$\rm{-0.1}$\% to 0.1\% in the range 20--70~\GeV. 
The corresponding systematic error is 
43 \MeV\ for electrons and 33 \MeV\ for muons. 

The systematic error due to the uncertainties on the resolution
is studied with Monte Carlo samples by comparing the fitted W mass 
with the nominal resolution to that with resolutions changed by 1.10 
for electrons (with the information of the electromagnetic calorimeter)
and 1.07 for muons (with the information of the central tracker). 
The corresponding errors are 43~\MeV\ and 12~\MeV\ respectively. 

\subsection{Background Treatment}

The background normalization is varied by $\pm$ 5\%, which corresponds 
to the error in the cross section measurement for 
the leptonic channel\cite{bib:cross-sec}. The resulting change in the 
fitted \mw\ is 11~\MeV.

\subsection{Parameterization of the Sensitive Variables}

A total of 152 parameters are obtained for  
the leptonic energy and the pseudo-mass considering 
center-of-mass energies from 183~\GeV\ to 207~\GeV\ for
all the defined classes. 
From these, 56~parameters depend on the mass of the W boson.
To check the systematic error associated with the parameterization
of the leptonic energy and the pseudo-mass distributions 
each parameter is varied independently by $\rm{\pm 1\sigma}$. 
This is repeated for all center-of-mass energies and all classes defined
for the leptonic energy and for the pseudo-mass.
Adding the resulting changes in quadrature yields a total systematic 
uncertainty due to the parameterization of 51~\MeV.

Since the  parameterization of the pseudo-mass does not describe a 
possible fall of the distribution above 85~\GeV\ we tried an alternative
parameterization by multiplying the standard parameterization by a linear 
function. This changes the fit result by 4~\MeV.

\subsection{Four-Fermion Effects}
Using events generated according to the CC03 diagrams instead of the
full set of four-fermion diagrams results in a mass bias of 24 \MeV.
This has been determined by reweighting events produced 
with \KoralW\ to the prediction of the CC03 matrix element.
The fit result has been corrected for the 
mass bias due to the use of the CC03 Monte Carlo at
center-of-mass of 183~\GeV\ and 189~\GeV.
The remaining uncertainty due to the modeling of the four fermion
final state is expected to be significantly smaller and has been
neglected in the estimation of the total systematic uncertainty.

\section{Outlook for Future Experiments}
This analysis has demonstrated for the first time that the 
correlation between the two leptons can be used as additional 
information to the single lepton energy spectra in the determination
of the W mass from leptonic W-pair decays. 
Due to the small correlation between the mass determination from
the pseudo-mass and from the lepton energy the uncertainty is reduced significantly
by combining the two measurements.
The systematic uncertainty could be significantly reduced in future
high-statistics experiments. The  uncertainties on the
energy scale, resolution and linearity are determined with data and
scale with the available statistics.
The error due to the parameterization depends on the number of available
Monte Carlo events and can be kept small by producing sufficiently large 
samples.

The different sensitivities of the
edges of lepton energy distributions and the pseudo-mass to the beam
energy result in a reduced systematic error from the uncertainty in the
beam energy in this analysis
compared to the \qqln\ and \qqqq\ channels \cite{bib:mass_189}. 
This will be important at future colliders, where the beam energy
measurements will be one of the limiting systematic uncertainties.

For the \qqln~ and \qqqq~ channels the hadronisation of the quarks 
gives rise to an important systematic uncertainty
which can only be estimated by the comparison of different hadronisation
models. The determination of the W mass in purely leptonic
W decays is free from this uncertainty.
As discussed, the total error in this channel should be competitive
with the \qqln\ and \qqqq\ channels with sufficiently high statistics.

\section{Summary}

The energy of the leptons produced in the reaction
\ee\ $\to$ \lnln\ and the pseudo-mass, a variable which is based on an
approximate kinematic reconstruction of the event, are used to measure
the mass of the W boson. Both variables are combined in a simultaneous
unbinned maximum likelihood fit, using data collected at center-of-mass
energies between 183~\GeV\ and 207~\GeV. The following result is obtained:
\begin{center}{$\rm{\mw\ = 80.41\pm 0.41\pm 0.13~\GeV,}$}
\end{center}
\indent where the first uncertainty quoted is statistical 
and the second is systematic. The result obtained is consistent with 
previous measurements of \mw\ \cite{bib:mass_189, bib:pdg00}.

\section{Acknowledgements}

We particularly wish to thank the SL Division for the efficient operation
of the LEP accelerator at all energies
 and for their close cooperation with
our experimental group.  We thank our colleagues from CEA, DAPNIA/SPP,
CE-Saclay for their efforts over the years on the time-of-flight and
trigger
systems which we continue to use.  In addition to the support staff at our
own
institutions we are pleased to acknowledge the  \\
Department of Energy, USA, \\
National Science Foundation, USA, \\
Particle Physics and Astronomy Research Council, UK, \\
Natural Sciences and Engineering Research Council, Canada, \\
Israel Science Foundation, administered by the Israel
Academy of Science and Humanities, \\
Minerva Gesellschaft, \\
Benoziyo Center for High Energy Physics,\\
Japanese Ministry of Education, Science and Culture (the
Monbusho) and a grant under the Monbusho International
Science Research Program,\\
Japanese Society for the Promotion of Science (JSPS),\\
German Israeli Bi-national Science Foundation (GIF), \\
Bundesministerium f\"ur Bildung und Forschung, Germany, \\
National Research Council of Canada, \\
Research Corporation, USA,\\
Hungarian Foundation for Scientific Research, OTKA T-029328, 
T023793 and OTKA F-023259,\\
Fund for Scientific Research, Flanders, F.W.O.-Vlaanderen, Belgium.\\



\newpage

\begin{table}[H]
\centering
{\begin{tabular}{|c|c|c|}
\hline
Event  &  Efficiency (\%) & Purity (\%) \\
\hline
$e\nu_{e}e\bar{\nu}_{e}$ & 75.5 & 91.2 \\
$\mu\nu_{\mu}\mu\bar{\nu}_{\mu}$ & 80.4 & 90.9 \\
$e\nu_{e}\mu\bar{\nu}_{\mu}$ & 77.8 & 95.9 \\
$e\nu_{e}\tau\bar{\nu}_{\tau}$ & 60.3 & 71.9 \\
$\mu\nu_{\mu}\tau\bar{\nu}_{\tau}$ & 60.6 & 75.5 \\
\hline
\end{tabular}
\caption{Signal efficiencies and purities for the leptonic events used in the analysis.}
\label{t:eff}}
\end{table}

\begin{table}[H]
\centering
{\begin{tabular}{|c|c|c|c|c|}
\hline
Center-of-mass & Integrated & Number of & Number of leptons & Number of events\\
energy (\GeV) & Luminosity ($\rm{pb^{-1}}$) & \lnln & ---leptonic energy--- & ---pseudo-mass---\\ 
              &                             & events &                      &  \\
\hline
183 & 57 & 71 & 77 & 26\\
189 & 183 & 278 & 309 & 80\\
192 & 29 & 52 & 51 & 12\\
196 & 77 & 144 & 169 & 32\\
200 & 74 & 134 & 144 & 25\\
202 & 37 & 82 & 87 & 16\\
205 & 82 & 129 & 141 & 14\\
207 & 128 & 211 & 248 & 29\\
\hline
All & 667 & 1101 & 1226 & 234\\
\hline
\end{tabular}
\caption{Integrated luminosities for the data from 183 to 207 GeV. 
The total number of events selected as \lnln, the number of leptons used for
the leptonic energy and the number of events used for the pseudo-mass
are shown.}
\label{t:events}}
\end{table}

\begin{table}[H]
\centering
{\begin{tabular}{c|c c c}
   &  $e$ & $\mu$ & $\tau$\\
\hline
     & & &\\
$e$  &  $LE/PM$ & $LE/PM$ & $LE$\\
$\mu$ & $LE/PM$ & $LE/PM$ & $LE$\\
$\tau$ & $LE$ & $LE$ & --\\
\end{tabular}
\caption{Classification of events used for the leptonic energy ($LE$) and the
pseudo-mass ($PM$) analysis.} 
\label{t:classi}}
\end{table}

\begin{table}[H]
\centering
{\begin{tabular}{|c|c|}
\hline
Systematic errors & Error (\GeV)\\ 
\hline
Beam energy & 0.011\\
Spread in the beam energy & 0.003\\
QED Corrections & 0.007 \\
Electromagnetic calorimeter scale  & 0.085  \\
Electromagnetic calorimeter resolution  & 0.043   \\
Electromagnetic calorimeter linearity & 0.043   \\
Central tracker scale  & 0.032  \\
Central tracker resolution  & 0.012 \\
Central tracker linearity & 0.033   \\
Background & 0.011 \\
Parameterization & 0.051 \\
\hline
Total & 0.127\\
\hline
\end{tabular}
\caption{Summary of systematic uncertainties on the \mw\ measurement.} 
\label{t:sys_un}}
\end{table}
\newpage

\newpage

\newpage

\vspace{-1cm}

\newpage
\vspace*{-2.5truecm}

\newpage
\newpage

\end{document}